\documentclass[10pt,aps,prb,reprint,amsfonts,amssymb,amsmath,floatfix,
superscriptaddress,twocolumn,
longbibliography,
]{revtex4-2}
\usepackage{microtype}
\usepackage{lmodern}
\usepackage[utf8]{inputenc}
\usepackage[T1]{fontenc}
\usepackage[usenames,dvipsnames]{xcolor}
\usepackage{graphicx}
\usepackage{hyperref}
\hypersetup{
    colorlinks=true,
    linkcolor=blue,
    citecolor=blue,   
    urlcolor=blue
}
\begin{document}

\title{Surface charge-transfer doping a quantum-confined silver monolayer beneath epitaxial graphene}

\author{Philipp Rosenzweig}
\email{p.rosenzweig@fkf.mpg.de}
\author{Hrag Karakachian}
\affiliation{Max-Planck-Institut f{\"u}r Festk{\"o}rperforschung, Heisenbergstraße 1, 70569 Stuttgart, Germany}
\author{Dmitry Marchenko}
\affiliation{Helmholtz-Zentrum Berlin f{\"u}r Materialien und Energie, Elektronenspeicherring BESSY II, Albert-Einstein-Straße 15, 12489 Berlin, Germany}
\author{Ulrich Starke}
\affiliation{Max-Planck-Institut f{\"u}r Festk{\"o}rperforschung, Heisenbergstraße 1, 70569 Stuttgart, Germany}

\date{\today}

\begin{abstract}
Recently the graphene/SiC interface has emerged as a versatile platform for the epitaxy of otherwise unstable, monoelemental, two-dimensional (2D) layers via intercalation. Intrinsically capped into a van der Waals heterostructure with overhead graphene, they compose a new class of quantum materials with striking properties contrasting their parent bulk crystals. Intercalated silver presents a prototypical example where 2D quantum confinement and inversion symmetry breaking entail a metal-to-semiconductor transition. However, little is known about the associated unoccupied states, and control of the Fermi level position across the bandgap would be desirable. Here, we $n$-type dope a graphene/2D-Ag/SiC heterostack via \emph{in situ} potassium deposition and probe its band structure by means of synchrotron-based angle-resolved photoelectron spectroscopy. While the induced carrier densities on the order of $10^{14}$ cm$^{-2}$ are not yet sufficient to reach the onset of the silver conduction band, the band alignment of graphene changes relative to the rigidly shifting Ag valence band and substrate core levels. We further demonstrate an ordered potassium adlayer ($2\times 2$ relative to graphene) with free-electron-like dispersion, suppressing plasmaron quasiparticles in graphene via enhanced metalization of the heterostack. Our results establish surface charge-transfer doping as an efficient handle to modify band alignment and electronic properties of a van der Waals heterostructure assembled from graphene and a novel type of monolayered quantum material.
\end{abstract}

\maketitle

\section{Introduction}
\label{sec:intro}
The experimental discovery of graphene \cite{novoselov2004,berger2004} can be considered a first avenue into the diverse and ever-emerging manifold of two-dimensional (2D) materials \cite{mas2011,bhimanapati2015}. Triggered by dimensional confinement, these 2D systems develop fascinating quantum properties that are absent from their bulk analogues, such as massless Dirac fermions in graphene \cite{novoselov2005} or valley-coupled giant spin splitting in transition-metal dichalcogenide monolayers \cite{zhu2011,xiao2012}. Going beyond individual atomic layers, their stacking into van der Waals (vdW) heterostructures even paves a way towards constructing designer quantum materials \cite{geim2013,novoselov2016}. Owing to weak vdW interlayer coupling, individual 2D sheets retain their basic structure and chemistry while the combined heterostructure gives rise to, e.g., expanded and tunable (opto)electronic \cite{dean2010,liu2016} or spintronic \cite{song2018,sierra2021} properties.

Boosted by this outstanding potential, the synthesis of 2D materials has come a long way in the last decade, not least with sample quality and reproducibility reaching remarkable degrees. This is exemplified by epitaxial graphene, today grown on wafer-scale SiC substrates with superb layer-thickness precision and almost defect free \cite{emtsev2009,kruskopf2016}. Such samples have further proven an ideal template for intercalation, i.e., the insertion of foreign atomic species at the graphene/SiC interface. Depending on the intercalated element, the electronic structure of graphene can be tailored so as to demonstrate maximal decoupling from the substrate \cite{riedl2009,sforzini2015}, superpotential-induced mini Dirac cones \cite{forti2016}, ambipolar doping levels \cite{emtsev2011,karakachian2022}, or extreme carrier densities up to \cite{mcchesney2010,rosenzweig2019,link2019} and even beyond Van Hove filling \cite{rosenzweig2020b}.

Yet, in view of the aforementioned designer quantum materials, the impact of intercalation reaches well beyond the sole manipulation of the overhead graphene layer---a fact that had largely escaped notice. Only recently, certain intercalated metals were found to order as monolayers on a triangular lattice in $1\times1$ epitaxial relationship to the SiC substrate, thereby giving rise to distinct band structures \cite{hayashi2017,yaji2019,forti2020,rosenzweig2020,briggs2020}. Besides protectively capping the ordered intercalant sheet \cite{briggs2019}, the graphene overlay retains its quasi-free-standing character and a 2D vdW heterostructure is readily formed \cite{forti2020,rosenzweig2020}. In contrast to conventional vdW heterostacks constructed from individually stable 2D sheets, only vertical confinement via the overhead graphene carpet will stabilize the atomically thin, triangular-lattice metals on the mesoscale. Consequently, intercalation or 'confinement heteroepitaxy' \cite{briggs2020} gives rise to a class of unique 2D metals with appealing quantum effects such as coexisting distinct types of spin splitting \cite{yaji2019} or superconductivity \cite{briggs2020}. Intriguingly, upon dimensional reduction at the graphene/SiC heterointerface, the prototype $sp$ metals Au \cite{forti2020} and Ag \cite{rosenzweig2020} even develop semiconducting electronic structures with the Fermi level pinned well above the valence band maximum in the latter case. Indeed, density functional theory (DFT) reproduces the emergence of a global, indirect band gap of 2D-Ag on SiC, driven by inversion symmetry breaking \cite{wang2020}. However, experimental access to its nonequilibrium band structure is so far scarce. A recent study could estimate the band gap size around $1$ eV via two-photon photoemission and found the effective mass of the conduction band minimum at $\overline{\Gamma}$ to deviate from \emph{GW}-corrected DFT \cite{lee2022}. This could point towards emerging many-body interactions with Dirac fermions in graphene, once again reinforcing the quantum behavior of 2D confined silver. In order to fully understand, manipulate and potentially exploit such interactions, controlling the position of the Fermi level throughout the semiconducting gap is essential. Eventually, pushing the Fermi level into the Ag conduction band upon $n$-type doping could further help to clarify its dispersion (and potential renormalizations) throughout an extended range in $k$-space by means of steady-state angle-resolved photoelectron spectroscopy (ARPES).

In this work, we $n$-type dope a graphene/2D-Ag/SiC heterostack to an electron density above $10^{14}$ cm$^{-2}$ via surface charge-transfer from adsorbed potassium and probe its band structure with ARPES. The graphene $\pi$ bands undergo a significant downshift, thereby changing their energetic alignment relative to the underlying valence band of monolayer Ag. The latter shifts rigidly with the Si $2p$ core levels, indicating a strong anchoring of interfacial silver to the SiC substrate. Despite substantial $n$-type doping, the Fermi level remains pinned inside the semiconducting gap of 2D-Ag and its conduction band minimum is not quite reached. Beyond acting as bare electron donor, the potassium adlayer forms an ordered superstructure as pinpointed by the emergence of a free-electron-like band at the Fermi level. The related metalization of the heterostructure is shown to efficiently suppress plasmaron excitation in the graphene layer. Our results highlight the potential of surface charge-transfer doping to control the electronic structure and quantum properties in a novel type of van der Waals heterostructure.

\section{Experiment}
\label{sec:exp}
On-axis, $n$-type doped 6H-SiC(0001) wafer pieces (SiCrystal GmbH) are used as primal substrates after chemical cleaning and atomic flattening via hydrogen etching \cite{ramachandran1998}. The vdW heterostructure of 2D silver and graphene is built up starting from the $(6\sqrt{3}\times6\sqrt{3})\mathrm{R}30^\circ$ carbon buffer layer reconstruction of 6H-SiC(0001). Inspired by Ref.~\cite{emtsev2009}, we grow this buffer layer \emph{ex situ} via Si sublimation at elevated temperatures around $1450^\circ$C in argon atmosphere ($800$ mbar). Silver is subsequently intercalated underneath in ultrahigh vacuum, turning the buffer layer into quasi-free-standing monolayer graphene. Concomitantly, the Ag atoms arrange on an atomically-thin triangular lattice that is epitaxial to SiC and hosts a distinct, semiconducting band structure as already described elsewhere \cite{rosenzweig2020}.

For the present study we have refined the \emph{in situ} intercalation protocol of Ag in comparison to our previous study \cite{rosenzweig2020}. Silver was first evaporated for $15$ min (nominal rate $4$ \AA/min) on the $6\sqrt{3}$ buffer layer heated to $400^\circ$C, followed by post-annealing to $600^\circ$C. In a second step, about $6$ nm of Ag were deposited at room temperature and the sample was finally heated to $700^\circ$C for $>1$ h. During sample preparation the temperature was measured by an infrared pyrometer (Impac IGA 140 series) assuming an emissivity of $63$\%. Samples were mounted on flag style holders (made from Mo) and heated indirectly.

The two-step preparation procedure yields sharper graphene and silver bands in comparison to Ref.~\cite{rosenzweig2020}. At the same time, the intrinsic $n$-type doping of the graphene/Ag heterostack turns out slightly lower (cf.\ above). Note that while the first step intercalates only small fractions of the surface as judged from low-energy electron diffraction, it still seems to open up intercalation pathways, presumably near SiC step edges \cite{chen2020} or by inducing a small number of defects in the graphene layer~\cite{briggs2019,briggs2020,liu2021,kotsakidis2021}. It is only by means of such pretreated samples that silver can succesfully be intercalated during the second preparation step and the resulting band structure displays optimal quality.

After confirming their integrity in the home lab, the as-prepared samples were taken to the synchrotron endstation in an ultrahigh vacuum transport suitcase (Ferrovac GmbH) ensuring a base pressure below $10^{-9}$ mbar.

ARPES was carried out using linearly polarized synchrotron radiation at the $1^2$ experimental station of the UE112 beamline at BESSY II, Helmholtz-Zentrum Berlin~\cite{helmholtz2018}. A hemispherical spectrometer with 2D electron detector (Scienta R8000) providing an ultimate angular resolution of $0.1^\circ$ was employed. The maximally achieved energy resolution of the entire setup (spectrometer and beamline) was $20$ meV. During all measurements, the base pressure was below $2\times10^{-10}$ mbar and unless explicitly stated otherwise, the sample was cooled to a temperature of $35$ K.

Surface charge-transfer doping was achieved via \emph{in situ} potassium (K) deposition on the cold sample from a commercial alkali metal dispenser (SAES Getters), operated at $5.6$ A, $1.3$ V as described elsewhere~\cite{rosenzweig2020b}. For a total deposition time of $7$ min, the density of adsorbed K atoms can be estimated at around $5\times10^{14}$ cm$^{-2}$, i.e., close to complete monolayer coverage according to the observed $2\times 2$ superstructure with respect to graphene (cf.\ Sec.\ \ref{subsec:korder}).

\section{Results and discussion}
\label{sec:results}
\subsection{Electronic structure of the pristine heterostack}
\label{subsec:pristine}
\begin{figure}[t]
	\centering
	\includegraphics{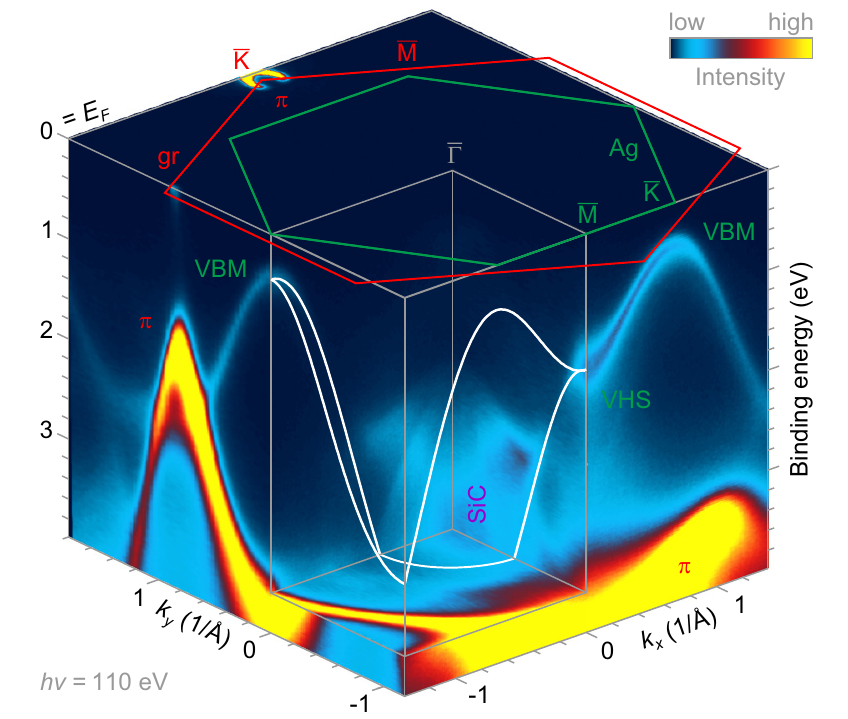}
	\caption{3D ARPES overview of the graphene/Ag/SiC heterostack (data collected at room temperature for $k_y<0$ and symmetrized for visualization). The 2D BZs of graphene (red) and silver (green) are indicated. Intense graphene $\pi$ bands superimpose the Ag valence band which hybridizes with the SiC bulk electronic structure towards $\overline{\Gamma}$. White curves represent the tight-binding model of Ref.~\cite{rosenzweig2020}, shifted in energy to match the Ag VBM and VHS at $\overline{\mathrm{K}}_\mathrm{Ag}$ and $\overline{\mathrm{M}}_\mathrm{Ag}$ respectively.}
	\label{fig1}
\end{figure}

Fig.~\ref{fig1} presents a 3D band structure overview of the graphene/2D-Ag/SiC heterostructure measured at a photon energy of $hv=110$ eV at room temperature. The intense $\pi$ bands of moderately $n$-doped quasi-free-standing monolayer graphene dominate the spectral intensity. Additional bands are evident that belong to intercalated Ag and disperse through a saddle-point Van Hove singularity (VHS) at $\overline{\mathrm{M}}_\mathrm{Ag}$ until peaking in a valence band maximum (VBM) at $\overline{\mathrm{K}}_\mathrm{Ag}$ well below the Fermi level $E_F$ \cite{rosenzweig2020,lee2022}. The associated 2D Brillouin zone (BZ), indicated by the green overlay on the Fermi surface, is rotated by $30^\circ$ relative to the one of graphene (red) and coincides with the surface BZ of SiC. This demonstrates the $1\times1$ epitaxial alignment of intercalated Ag in a triangular lattice on the Si-terminated substrate. White curves represent the 2D $s$-orbital tight-binding model of Ref.~\cite{rosenzweig2020} that traces the Ag valence band remarkably well in the vicinity of the BZ border while, by definition, it cannot capture the obvious hybridization with SiC bulk bands around $\overline{\Gamma}$. At the same time, matrix element effects suppress the spectral weight of 2D-Ag within intermediate momentum ranges along $\overline{\Gamma\mathrm{K}}_\mathrm{Ag}$ ($k_y=0$) and $\overline{\Gamma\mathrm{M}}_\mathrm{Ag}$ ($k_x=0$) inside the first BZ. Note a rigid downshift of $E_F$ by $\approx0.1$ eV (i.e., reduced intrinsic $n$ doping) as compared to our previous study \cite{rosenzweig2020} which reproducibly results from our optimized intercalation procedure (cf.\ Sec.\ \ref{sec:exp}). Yet, the relative dispersion of the Ag valence band probed at $hv=110$ eV remains consistent with the tight-binding model derived from ARPES at significantly lower photon energies. As such, the Ag-related dispersion is shown to be independent of $h\nu$ over a wide range, reinforcing its 2D nature.

\begin{figure*}[t]
	\centering
	\includegraphics{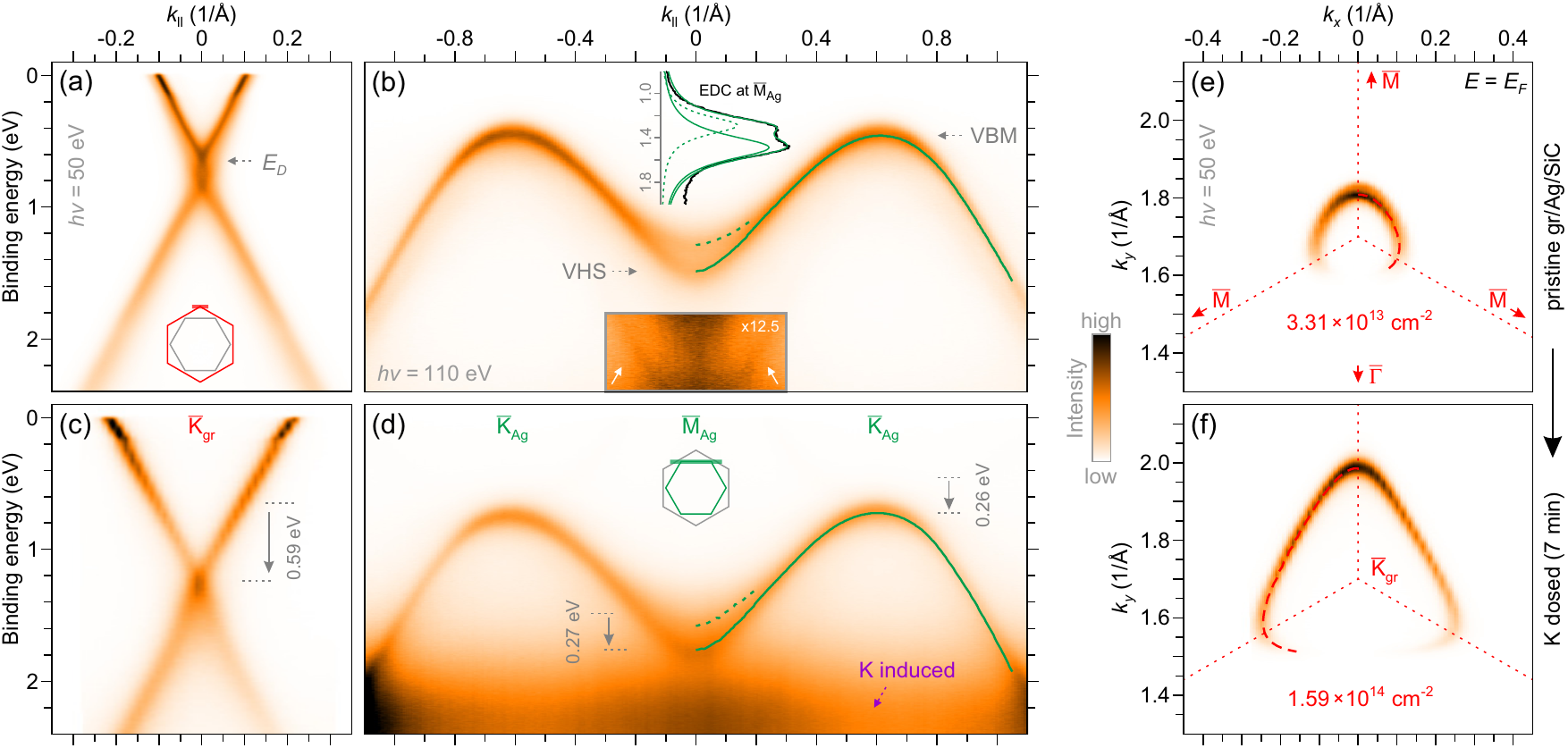}
	\caption{(a), (b) Energy-momentum cuts for pristine graphene/Ag/SiC slicing (a) the graphene Dirac cone at $\overline{\mathrm{K}}_\mathrm{gr}$, perpendicular to $\overline{\Gamma\mathrm{K}}_\mathrm{gr}$ and (b) the Ag valence band along the $\overline{\mathrm{KMK}}$ border of its BZ. The VHS at $\overline{\mathrm{M}}_\mathrm{Ag}$ splits up into two peaks (fitted EDC in the top inset). Enhanced contrast at the bottom center of (b) reveals replica bands of Ag (white arrows) backfolded via graphene reciprocal lattice vectors. (c), (d) Same as (a), (b) after $7$ min of potassium deposition. Owing to $n$-type charge-transfer doping, the Dirac point binding energy $E_D$ increases by $0.59$ eV while the Ag valence band undergoes a rigid shift of only $0.26$ eV. Concomitantly, the elongated region around the Dirac point is reshaped (cf.\ Fig.~\ref{fig6}). Additional spectral weight appearing below $2$ eV in (d) can be attributed to ionized potassium~\cite{koch2018}. (e), (f) The carrier density of graphene increases by $1.26\times10^{14}$ cm$^{-2}$ as determined from the area enclosed by its Fermi surface pockets (e) before and (f) after potassium deposition.}
	\label{fig2}
\end{figure*}

Fig.~\ref{fig2}(a) displays a high-resolution, low-temperature ($35$ K) energy-momentum cut through the Dirac cone of quasi-free-standing monolayer graphene centered at $\overline{\mathrm{K}}_\mathrm{gr}$, directed perpendicular to $\overline{\Gamma\mathrm{K}}_\mathrm{gr}$ (cf.\ inset). The crossing of the $\pi$ band branches is found to severly renormalize into a diamond-like feature, indicative of plasmaron quasiparticle formation \cite{polini2008,bostwick2010,walter2011} and analyzed in detail within the scope of Fig.~\ref{fig6} below. For later reference, we determine the Dirac point $E_D$ at $0.65\pm0.02$ eV below $E_F$ as per the crossing of the upper branches (linearly extrapolated from within the binding energy window $0.2$--$0.5$ eV).

A detailed view of the silver-related valence band dispersion along the $\overline{\mathrm{KMK}}$ border of the associated BZ is given by the ARPES cut of Fig.~\ref{fig2}(b). Solid and dashed green curves overlayed for $k_\parallel\geq0$ trace the band course as per energy distribution curve (EDC) fitting. From a single VBM at $0.45\pm0.02$ eV below $E_F$, the dispersion is found to split up near $\overline{\mathrm{M}}_\mathrm{Ag}$ and eventually forms two distinct saddle-point VHSs with binding energies of $1.29$ and $1.49\pm0.03$ eV, respectively (cf.\ fitted EDC in the inset). Note that this detail is either absent or not resolved in the room-temperature data of Fig.~\ref{fig1}. Splitting of a VHS has hitherto been known from high-$T_c$ superconductors where it is possibly intertwined with structural instabilities \cite{labbe1987,barisic1987,friedel1987,markiewicz1997} or---more recently---from twisted bilayer graphene, driven by electronic correlations \cite{li2017,kerelsky2019,choi2019,liu2019}. In either of the above cases, the effect vanishes once the VHS departs from $E_F$. No analogy can thus be drawn to our system, not least because the energy separation of $\approx0.2$ eV exceeds these previously reported splittings by an order of magnitude. Since both VHS peaks share comparable spectral weights, we can exclude the presence of a distinct minority phase of intercalated silver with its own saddle-point binding energy at $\overline{\mathrm{M}}_\mathrm{Ag}$. Any such phase must rather occupy a substantial portion on the SiC surface without introducing a differential doping level in graphene [cf. the single, dominant set of $\pi$ bands in Fig.~\ref{fig2}(a)]. This is unlikely considering the general sensitivity of epitaxial graphene to the specific composition of the interface \cite{emtsev2011,rosenzweig2019,forti2020}. Enhanced contrast at the bottom center of Fig.~\ref{fig2}(b) reveals steep replica features (highlighted by white arrows) of the silver valence band originating from final-state photoelectron diffraction at the graphene lattice. Yet, such replicas are extremely weak and the available reciprocal lattice vectors are not suitable to generate an apparent second VHS along the $\overline{\mathrm{KMK}}$ border of the first BZ of 2D-Ag via backfolding. It could instead be speculated as to whether spin-orbit coupling underlies the splitting of the VHS, e.g., in analogy to the split-off $j=1/2$ valence band present at the BZ center of (bulk) group IV and III-V semiconductors \cite{yu-book}. In contrast to the spin-split valence band maximum at the corners of the hexagonal BZ of monolayer transition-metal dichalcogenides \cite{zhu2011,xiao2012}, exchange-type splitting of the VHS of 2D-Ag at $\overline{\mathrm{M}}$ into two spin-polarized counterparts is hardly conceivable: lifting the spin degeneracy at the time-reversal invariant $\overline{\mathrm{M}}$ points \cite{fu2007} would require the breaking of time-reversal symmetry. In any case, none of the DFT calculations available for the graphene/2D-Ag/SiC system \cite{hsu2012,wang2020,lee2022} exhibit a feature reminiscent of the observed VHS splitting at $\overline{\mathrm{M}}_\mathrm{Ag}$ (not even when spin-orbit coupling is included) and the origin of this effect remains an open question to be addressed in future studies.

\subsection{Response to potassium deposition}
\label{subsec:response}
In order to manipulate the Fermi-level pinning within the graphene/Ag/SiC heterostack, we have deposited potassium (K) \emph{in situ} at low temperature ($35$ K) for $7$ min (cf.\ Sec.\ \ref{sec:exp}). As demonstrated in the energy-momentum cut of Fig.~\ref{fig2}(c), electron transfer from adsorbed K entails a downshift of the Dirac point of graphene by $\approx0.59$ eV to $E_D=1.24\pm0.02$ eV (again determined via linear extrapolation of the upper branches within $0.6$--$1.0$ eV, cf.\ above) while the Dirac cone is concurrently reshaped. On the one hand, the dispersion acquires a prominent kink at $\approx0.2$ eV below $E_F$ due to an enhancement of electron-phonon coupling \cite{bostwick2007,calandra2007,pletikosic2012}. On the other hand and more striking, the elongated crossing region owing to electron-plasmon coupling transforms back into a singular Dirac point. The above coupling mechanisms and their modification induced by potassium doping are discussed in greater detail in the context of Figs.~\ref{fig5} and \ref{fig6} below.

The $\overline{\mathrm{KMK}}$-dispersion of 2D confined silver is in turn shown in Fig.~\ref{fig2}(d). Upon $7$ min of potassium exposure, additional spectral background appears for binding energies $\gtrsim2$ eV due to a lower-lying nondispersive state, which can be assigned to ionized K atoms on the sample surface \cite{koch2018}. The VBM and VHS of 2D silver shift to higher binding energies by $0.26$ and $0.27\pm0.02$ eV, respectively. At the same time, the VHS splitting of $\approx0.2$ eV persists and the relative band course remains unaltered (cf.\ solid and dashed green curves). Note that this rigid valence band shift amounts to less than half as much in comparison to the displacement of $E_D$. Potassium doping is thus shown to manipulate the band alignment within the metal/semiconductor heterostack, shifting the Dirac point energy of graphene away from the Ag VBM and towards the (split) VHS. However, the K-induced binding energy shifts are not yet sufficient for any signature of the silver-related conduction band to appear along $\overline{\mathrm{KMK}}_\mathrm{Ag}$.

\begin{figure*}[t]
	\centering
	\includegraphics{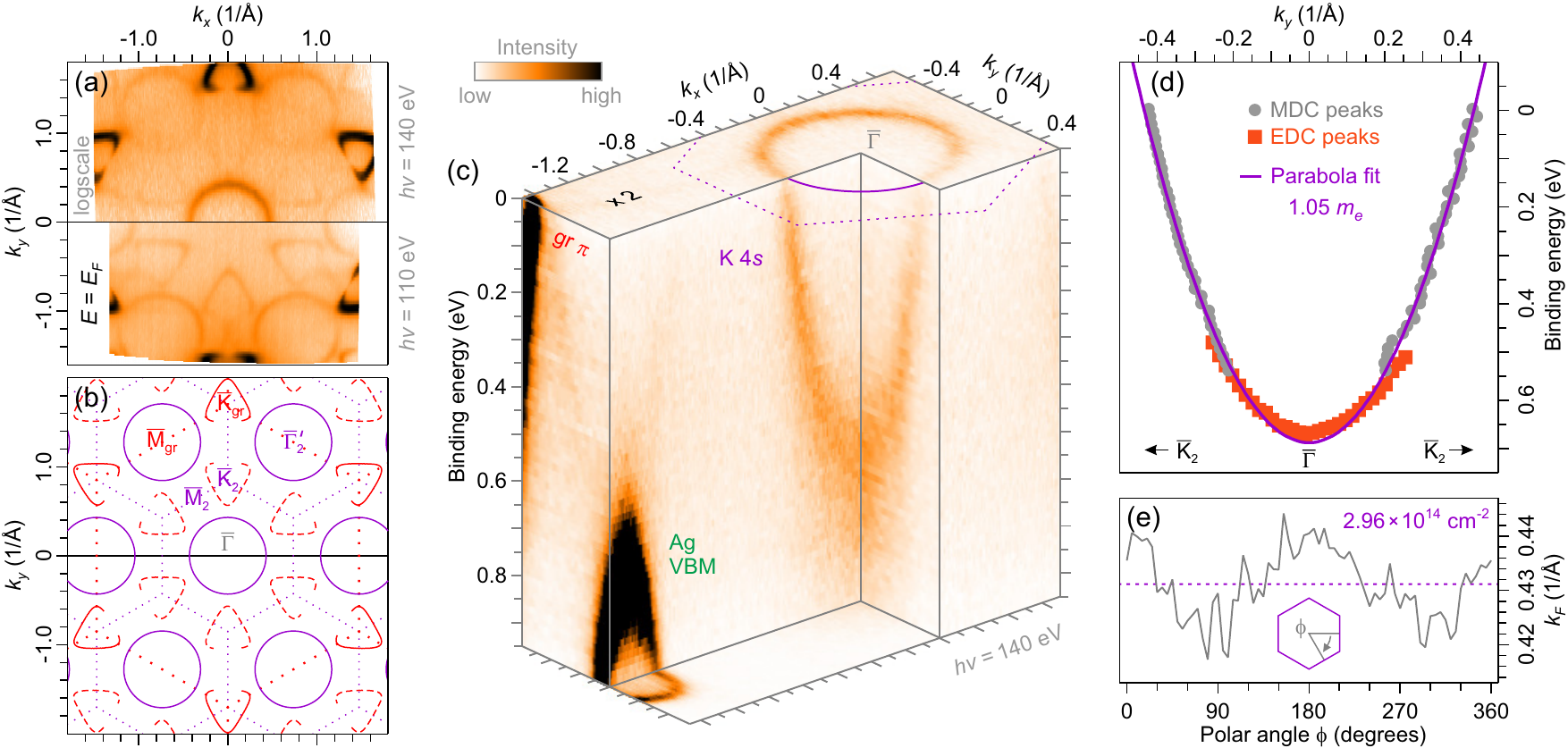}
	\caption{(a) Experimental Fermi surface after potassium deposition probed over a wide range of parallel momenta at $h\nu=110$ eV (bottom) and $140$ eV (top). The latter dataset was obtained for $k_y<0$ and mirrored along $k_y=0$ for visualization purposes. Besides the intense $\pi^*$ pockets around $\overline{\mathrm{K}}_\mathrm{gr}$, a potassium-induced circular contour around $\overline{\Gamma}$ as well as graphene replicas emerge, both repeating with a $2\times2$ periodicity relative to the graphene lattice. (b) Schematized Fermi surface highlighting the original $\pi^*$ pockets (solid red) with their $2\times2$ replicas (dashed red) and the circular contours (solid purple). $1\times1$ (dotted red) and $2\times2$ (dotted purple) BZs of graphene are also indicated. (c) 3D view of the potassium-induced band at $\overline{\Gamma}$. The dispersion is consistent with the free-electron-like expectation for K $4s$. (d) Parabola fit (purple curve) to the spectral MDC (grey circles) and EDC maxima (orange squares). (e) Extracted Fermi wave vector $k_F$ as a function of polar angle $\phi$ along the pocket. $k_F$ is essentially constant (dashed purple line shows the average value) and $2.96\times10^{14}$ electrons/cm$^2$ remain within the $2\times2$-K adlayer.}
	\label{fig3}
\end{figure*}

While the band shifts provide a qualitative picture of charge transfer, the carrier density transferred onto graphene can be accurately quantified based on the evolution of its Fermi surface. Figs.~\ref{fig2}(e) and (f) display the electron Fermi pocket of the $\pi^*$ band around $\overline{\mathrm{K}}_\mathrm{gr}$ before and after potassium deposition, respectively. According to Luttinger's theorem \cite{luttinger1960,luttinger1960a} the respective carrier density is given by $n=A/\pi^2$, where $A$ is the area enclosed by the pocket and the presence of two such pockets in the BZ has been taken into account. We determine $A$ with a relative error below $5$\% by fitting radial momentum distribution curves (MDCs) in steps of $2^\circ$ and integrating the associated Fermi wave vector $k_F$ [red dashed curves in Figs.~\ref{fig2}(e) and (f)] along the pocket outside the first BZ. Data inside the first BZ are excluded from the analysis due to the well-known suppression of spectral weight along the $\overline{\Gamma\mathrm{K}}_\mathrm{gr}$ corridor resulting from sublattice interference \cite{shirley1995,gierz2011}. Hence, we find the $n$-type carrier density of pristine Ag-intercalated graphene to increase from $3.31\times10^{13}$ cm$^{-2}$ to $1.59\times10^{14}$ cm$^{-2}$ upon potassium deposition. This corresponds to a substantial charge transfer of $1.26\times10^{14}$ cm$^{-2}$ or $0.033$ electrons per graphene carbon atom.

\subsection{Ordering and free-electron-like metallicity of the potassium adlayer}
\label{subsec:korder}
Going beyond the evolution of bands inherent to the graphene/Ag heterostructure, we now examine how the adsorbed potassium contributes itself to the electronic structure apart from bare charge transfer. Fig.~\ref{fig3}(a) shows the Fermi surface of the K-doped heterostack probed over a wide momentum range at $hv=110$ eV (bottom) and $140$ eV (top). A large circular pocket around $\overline{\Gamma}$ can be observed for $hv=140$ eV (suppressed for $110$ eV). Its repeated contours at $\overline{\mathrm{M}}_\mathrm{gr}$ are visible for both photon energies and match a $2\times2$ superperiodicity relative to the graphene lattice. Replicated $\pi^*$ pockets now appearing at the $\overline{\mathrm{K}}_2$ points of the associated $2\times2$ BZ corroborate this scenario, illustrated in the Fermi surface schematic of Fig.~\ref{fig3}(b). The induced $2\times2$ superstructure can be assigned to an ordering of the adsorbed potassium atoms, which partially donate their $4s$ electron to the underlying heterostack, thus leaving behind a Fermi surface pocket at intermediate filling. The 3D ARPES intensity plot of Fig.~\ref{fig3}(c) provides a closeup view of the corresponding K $4s$ state centered at $\overline{\Gamma}$ ($hv=140$ eV). Starting from a band bottom at $0.67\pm0.02$ eV below $E_F$, its parabolic dispersion suggests a free-electron-like nature as expected for an alkali metal. Fig.~\ref{fig3}(d) shows the extracted EDC (orange squares) and MDC peaks (grey circles) of the K $4s$ band along the $\overline{\mathrm{K}\Gamma\mathrm{K}}$ direction of the $2\times2$ BZ. A parabolic fit yields an effective mass $m^*=1.05\pm0.05$ in units of the free electron mass $m_e$. As further shown in Fig.~\ref{fig3}(e), the Fermi wave vector $k_F=0.43\pm0.01$ {\AA}$^{-1}$ does not vary significantly along the entire pocket. This all confirms the almost ideal free-electron character of the metallic potassium overlayer whose residual carrier density can be directly extracted from the pocket's Luttinger area as $n=(2.96\pm0.15)\times10^{14}$ cm$^{-2}$.

It might have been tempting to instead assign the feature at $\overline{\Gamma}$ to the conduction band of 2D silver whose minimum is expected right there \cite{wang2020,lee2022}. However, since the Ag VBM now sits at a binding energy of $\approx0.7$ eV, very close to the bottom of the parabola [cf.\ Fig.~\ref{fig3}(c)], this would mean that the material's indirect band gap of $\approx 1$ eV has essentially closed by virtue of potassium deposition. Although surface charge-transfer doping with K atoms has indeed been demonstrated to tune and even close band gaps in 2D semiconductors \cite{ohta2006,kim2015,chen2018}, this usually requires direct proximity to the ionized dopant atoms and concomitant renormalizations also occur in the respective valence bands. Both these factors do not apply here and the above hypothesis can thus be ruled out.

Note that $2\times2$ order of adsorbed or intercalated potassium and the presence of a metallic K $4s$ state is well known from bulk graphite \cite{bennich1999,algdal2006}. For monolayer graphene, a K-$2\times2$ overlayer with a free-electron parabola at $\overline{\Gamma}$ has only been demonstrated on a metallic Ir(111) substrate \cite{struzzi2016} (in overall good agreement to the present study) while corresponding data is not yet available for epitaxial graphene on SiC. Surprisingly, Yb-intercalated graphene on SiC \cite{rosenzweig2020b} does not display a $2\times2$ superstructure upon potassium adsorption, although analogous charge transfer could be achieved using the same experimental setup under similar conditions. Taking into account that the carrier density of graphene in Ref.~\cite{rosenzweig2020b} was already on the order of $3\times10^{14}$ cm$^{-2}$ even before K deposition, this suggests an important role of the graphene substrate itself in determining the structural and electronic properties of potassium on top. Going beyond bare charge transfer from and establishing metallic behavior in the overlayer is then again a key to influence the detailed electronic properties of the heterostack as discussed in connection with Fig.~\ref{fig6} below.

\subsection{Overview of charge transfer and band alignment}
\label{subsec:overview}
\begin{figure}[b]
	\centering
	\includegraphics{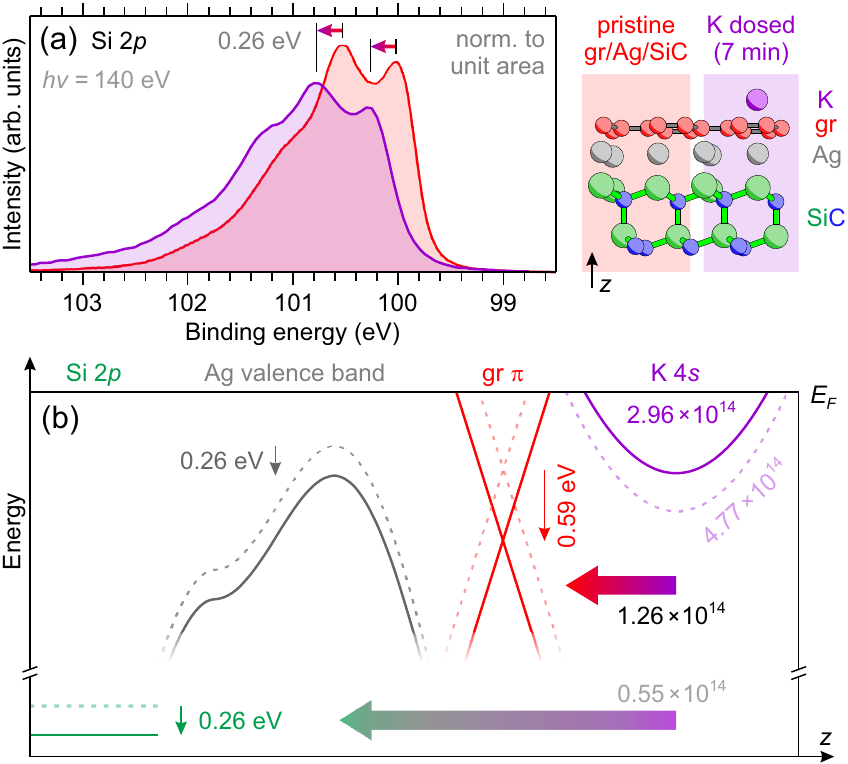}
	\caption{(a) Si $2p$ core level spectra (normalized to unit area) for the pristine sample (red) and after potassium deposition (purple). Si $2p$ core level and Ag valence band shift rigidly with each other (cf.\ Fig.~\ref{fig2}). (b) Changes in band alignment and ocurring charge transfers (electrons/cm$^2$) upon potassium deposition (schematic diagram). Dashed curves represent the pristine sample or a hypothetical, isolated layer of potassium not donating any charges, solid curves the $n$-type doped sample with $2\times2$-K on top. A ball-stick side view of the (K/)graphene/Ag/SiC heterostack is shown at the top right.}
	\label{fig4}
\end{figure}

Now that the electronic bands of the 2D graphene, silver and potassium layers have been addressed, it is instructive to also examine the response of the SiC substrate upon surface charge-transfer doping. To this end, Fig.~\ref{fig4}(a) displays the Si $2p$ core level for the pristine heterostack (red curve) where the contributions of surface and bulk Si are comparable owing to a high surface sensitivity at the employed photon energy of $140$ eV. Distinct Si $2p$ doublets thus lead to a relatively broad spectrum where the surface component is located on the lower binding energy side due to interaction with interfacial Ag \cite{rosenzweig2020,forti2020}. After potassium deposition the entire spectrum shifts by $\approx 0.26$ eV and becomes slightly broader towards higher binding energies [purple curve in Fig.~\ref{fig4}(a)]. This increased asymmetry of the core-level spectrum could be explained through energy loss effects, now ocurring via excitations in the metallic potassium overlayer \cite{bennich1999,chis2014}. Most notably, the K-induced binding energy shifts of the Si $2p$ core level and the valence band of 2D-Ag [cf.~Fig.~\ref{fig2}(d)] are identical, i.e., the relative band alignment between the two semiconducting materials is preserved. This emphasizes the strong anchoring of interfacially confined silver to the SiC substrate in what could be considered a 2D surface silicide---yet with a largely self-standing Ag-related electronic structure, wide parts of which are contained inside the band gap of SiC. The uniform energetic shifts after potassium deposition can hence be attributed to a change in semiconductor band bending at the interface to metallic graphene in response to its increased $n$-type carrier density that exceeds $10^{14}$ cm$^{-2}$.
Consistent with the prototypical vdW interfaces of 2D transition metal dichalcogenides (TMDCs) on epitaxial graphene \cite{coydiaz2015,miwa2015,forti2017,nakamura2020}, the 2D silver/graphene heterojunction displays only weak vdW coupling in view of the quasi-free-standing nature of graphene's $\pi$ bands [cf.~Figs.~\ref{fig2}(a) and (c)]. On the other hand, $E_F$ turns out to be moderately pinned to the semiconducting gap of Ag with an estimated strength \cite{tung2014} $S\approx\Delta E_{\mathrm{VBM}}/\Delta E_D\approx0.45$. Here, $\Delta E_{D}$ ($\Delta E_{\mathrm{VBM}}$) denotes the binding energy shift of $E_D$ (the Ag VBM), the work function of the system is assumed to change rigidly with $E_D$, and band gap renormalizations (highly unlikely as discussed in the previous section) are not taken into account. This is markedly different from the reported Schottky-Mott behaviour of TMDC-on-graphene heterostructures where Fermi-level pinning is basically absent ($S\approx 1$) \cite{liu2016b,lequang2017}. Despite its semiconducting band structure, intercalated 2D silver must thus be clearly delimited from the well-established family of monolayer TMDC semiconductors that are routinely synthesized on top of epitaxial graphene.

The scheme of Fig.~\ref{fig4}(b) summarizes the valence band alignment of the heterostack before (dashed curves) and after potassium deposition (solid curves), indicating also the extracted charge transfers. A hypothetical, isolated monolayer of K atoms on a triangular lattice with twice the lattice constant of graphene should have its $4s$ band exactly half filled at $4.77\times10^{14}$ cm$^{-2}$. While we observe a residual carrier density of $2.96\times10^{14}$ in the $2\times2$ potassium adlayer, the one of graphene increases only by $1.26\times10^{14}$ cm$^{-2}$, i.e., there is an apparent charge leakage of $0.55\times10^{14}$ cm$^{-2}$. Note that already less than $5$\% of the dangling bonds of an unsaturated SiC surface could provide the necessary reservoir density of states around $E_F$ to absorb these charges. Such a low number of residual dangling bonds is conceivable and still consistent with a large-area, homogeneous intercalation as evidenced by the band structure quality in Figs.~\ref{fig1} and \ref{fig2}. At the same time, it could also help to explain the moderate Fermi-level pinning inside the semiconducting band gap of 2D-Ag without having to speculate about additional in-gap states forming at the silver/graphene interface upon intercalation.

Although the conduction band minimum of monolayer Ag is not quite reached, introducing $> 10^{14}$ electrons per cm$^{2}$ into the heterostack is still sufficient to switch the majority carrier type of 2D-Ag from $p$ ($E_F$ closer to the VBM) to $n$ (closer to the conduction band minimum) when taking into account the estimated gap size of Ref.~\cite{lee2022}. Extrapolating the observed band shifts to the even higher charge transfers that can in principle be achieved via more aggressive potassium deposition \cite{rosenzweig2020b}, a semiconductor-to-metal transition of 2D confined Ag could eventually become possible.

\subsection{Many-body interactions in graphene}
\label{subsec:interactions}
\begin{figure}[b]
	\centering
	\includegraphics{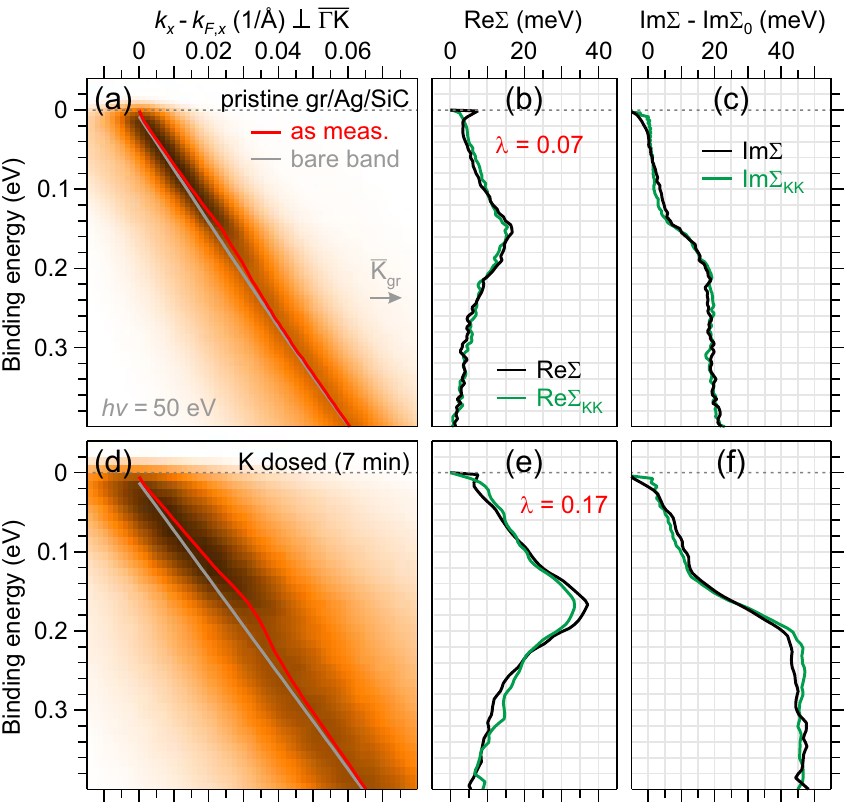}
	\caption{Self-consistent analysis of electron-phonon coupling in Ag-intercalated graphene (a)--(c) before and (d)--(f) after potassium deposition. (a), (d) ARPES cuts perpendicular to $\overline{\Gamma\mathrm{K}}_\mathrm{gr}$ (momentum given relative to Fermi-level crossing $k_{F,x}$) indicating the renormalized experimental dispersion obtained from MDC fitting (red) and the extracted bare band (grey). (b), (e) Real parts of the spectral function as per extracted bare band ($\mathfrak{Re}\Sigma$, black) and Kramers-Kronig transform of $\mathfrak{Im}\Sigma$ ($\mathfrak{Re}\Sigma_\mathrm{KK}$, green). (c), (f) Offset-corrected imaginary parts of the spectral function as per extracted bare band ($\mathfrak{Im}\Sigma$, black) and Kramers-Kronig transform of $\mathfrak{Re}\Sigma$ ($\mathfrak{Im}\Sigma_\mathrm{KK}$, green). The electron-phonon coupling constant $\lambda$ is estimated from a linear fit to $\mathfrak{Re}\Sigma$ within $20$--$60$ meV from $E_F$ and increases by a factor of $2.5$ through $n$-type charge-transfer doping.}
	\label{fig5}
\end{figure}

ARPES essentially probes the single-particle spectral function of a bare band dispersion that is subject to self-energy corrections \cite{damascelli2003}. The underlying many-body interactions can then be quantified based on a detailed analysis of the spectral function that disentangles the bare band and the imaginary self-energy $\Sigma=\mathfrak{Re}\Sigma+i\mathfrak{Im}\Sigma$ from the renormalized experimental dispersion. In this regard, electron-phonon coupling in graphene is understood to entail a kink in the $\pi$ band dispersion at $\approx 0.2$ eV below $E_F$ [cf.\ Figs.\ \ref{fig2}(a) and (c)] that has been studied extensively for different (intercalated) graphene systems \cite{bostwick2007,calandra2007,pletikosic2012,bisti2021}. Fig.\ \ref{fig5}(a) shows an ARPES energy-momentum cut perpendicular to $\overline{\Gamma\mathrm{K}}_\mathrm{gr}$, zooming in on the vicinity of the $\pi$-band Fermi-level crossing $k_{F,x}$ for the pristine graphene/Ag/SiC system. The red curve indicates the slightly kinked experimental dispersion as determined from Lorentzian MDC fitting. Employing the self-consistent algorithm of Ref.~\cite{pletikosic2012}, we have subsequently reconstructed the bare band dispersion [grey curve in Fig.\ \ref{fig5}(a)] and the real $\mathfrak{Re}\Sigma$ and imaginary parts $\mathfrak{Im}\Sigma$ of the self-energy as indicated by the black curve in Figs.\ \ref{fig5}(b) and (c), respectively. Since, by definition, $\mathfrak{Re}\Sigma$ and $\mathfrak{Im}\Sigma$ must satisfy causality, they should be interconnected by Kramers-Kronig (KK) relations, i.e., $\mathfrak{Im}\Sigma$ is the KK transform of $\mathfrak{Re}\Sigma$ and vice versa \cite{damascelli2003,pletikosic2012}. The hence determined functions $\mathfrak{Re}\Sigma_\mathrm{KK}$ and $\mathfrak{Im}\Sigma_\mathrm{KK}$ are indicated by the green curves in Figs.\ \ref{fig5}(b), (c). A very good agreement with their directly reconstructed counterparts $\mathfrak{Re}\Sigma$ and $\mathfrak{Im}\Sigma$ prooves the high degree of self-consistency of the present spectral function analysis. From the slope of $\mathfrak{Re}\Sigma$ right at $E_F$, the electron-phonon coupling strength $\lambda=-\mathrm{d}/\mathrm{d}E(\mathfrak{Re}\Sigma)\vert_{E_F}$ could, in principle, be directly determined. However, since finite-resolution effects are known to affect the precise dispersion and hence $\mathfrak{Re}\Sigma$ right around the Fermi level \cite{levy2014}, we instead estimate $\lambda=0.07\pm0.01$ from a linear fit to $\mathfrak{Re}\Sigma$ within the binding energy window of $20$--$60$ meV.

Figs.\ \ref{fig5}(d)--(f) present the analogous coupling analysis for the K dosed sample, where the kinked dispersion comes out much more prominent and we find $\lambda=0.17\pm0.02$. On the one hand, this value is well in line with K-adsorbed graphene on Ir(111) \cite{pletikosic2012} that shows a doping level similar to the present case. Assuming that the same phonon modes contribute to the overall coupling irrespective of the actual doping level, $\lambda$ is expected to be simply proportional to the density of states at $E_F$ \cite{calandra2007}, that is, to the square root $\sqrt{n}$ of graphene's carrier density. Indeed, $\lambda$ and $\sqrt{n}$ are found to increase almost equally upon potassium doping by factors of $2.4$ and $2.2$, respectively. The small difference between the two scaling factors can be attributed to a slight departure of graphene's $\pi$ band dispersion from linearity (i.e., velocity renormalizations) owing to its high carrier density $>10^{14}$ cm$^{-2}$. By scaling $\mathfrak{Re}\Sigma$ and $\mathfrak{Im}\Sigma$ of the pristine sample accordingly, both curves essentially coincide with their counterparts after potassium deposition. We thus conclude that the ordered K-$2\times2$ overlayer does not contribute extrinsic phonon modes to which the Dirac charge carriers of graphene couple in a significant manner. To the same extent, emerging couplings to other intrinsic (low-energy) phonon modes of graphene boosted by its enhanced carrier density can be ruled out. Note further that the coupling strength for the pristine graphene/Ag/SiC heterostack can also be extrapolated from, e.g., H-intercalated graphene on SiC \cite{forti2011} via the simple proportionality relation $\lambda\propto\sqrt{n}$. This suggests the absence of any measurable contribution from phonon modes that are specific to the ordered 2D-Ag interlayer only.

\begin{figure}[b]
	\centering
	\includegraphics{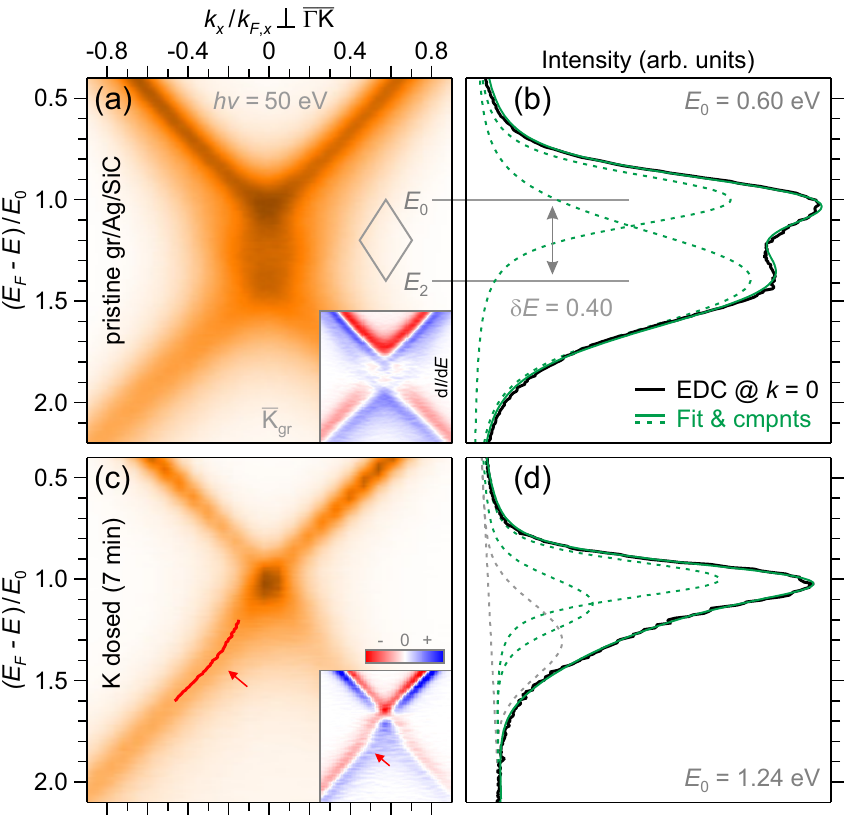}
	\caption{Electron-plasmon coupling in (a), (b) pristine and (c), (d) K-deposited graphene/Ag/SiC. (a), (c) ARPES cuts perpendicular to $\overline{\Gamma\mathrm{K}}_\mathrm{gr}$. Insets show the respective derivative along the energy axis. (b), (d) corresponding EDCs at $k=0$ (solid black) with their fits (solid and dashed green/grey). Energy and momentum axes are scaled to the upper Dirac point $E_0$ and the Fermi vector $k_{F,x}$, respectively. (a) The plasmaron-reshaped branch crossings (grey diamond, width $\delta E$) practically transform back into (c) a single Dirac point by introducing the metallic potassium overlayer (cf.\ Fig.~\ref{fig3}). Simultaneously, a kink develops at higher binding energy [red arrows and red curve tracking the MDC maxima in (c)].}
	\label{fig6}
\end{figure}

We now turn to the strong renormalization around graphene's Dirac point [cf.\ Fig.\ \ref{fig2}(a)]. A zoom-in ARPES cut of the respective region is shown in Fig.\ \ref{fig6}(a) where two Dirac points, i.e., a splitting into an upper and lower branch crossing, can clearly be discerned. Note that for evaluation and comparison purposes the binding energy and momentum ($k_x$) axes have been scaled relative to the upper Dirac point $E_0=0.60$ eV \footnote{Note the small difference in determining the upper Dirac crossing for the pristine sample as per EDC fitting at $\overline{\mathrm{K}}_\mathrm{gr}$ ($E_0=0.60$ eV, Sec.\ \ref{subsec:interactions}) and via linear extrapolation of the $\pi$-band branches ($E_D=0.65$ eV, Sec.\ \ref{subsec:pristine}). This deviation---not affecting any of the conclusions drawn in this work---can be attributed to the strong renormalization of the conical dispersion in closest vicinity of $\overline{\mathrm{K}}_\mathrm{gr}$. In fact, $E_0=E_D=1.24$ eV turn out to be identical after $n$-type doping when the renormalization is much less pronounced.} and the Fermi wave vector $k_{F,x}=0.10$ {\AA}$^{-1}$, respectively \cite{bostwick2010,walter2011}. The peculiar shape of the Dirac cone can be understood to result from coupling between the elementary Dirac fermions and plasmons in graphene, thereby forming composite plasmaron quasiparticles \cite{polini2008,bostwick2010,walter2011}. In consequence, a plasmaron-dressed band emerges that is intertwined with the bare Dirac cone. This gives rise to a diamond-like shape around $\overline{\mathrm{K}}_\mathrm{gr}$ as indicated by the grey overlay and highlighted by the spectral derivative along the energy axis in the inset of Fig.\ \ref{fig6}(a). Following established procedures \cite{bostwick2010,walter2011}, the effective coupling constant $\alpha_G$ of graphene can be estimated from the scaled energy width $\delta E=\vert(E_2-E_0)/E_0\vert$ of the plasmaron diamond. Fitting the EDC at $\overline{\mathrm{K}}_\mathrm{gr}$ ($k=0$) with two peaks corresponding to the upper ($E_0$) and lower Dirac point ($E_2$) as shown in Fig.\ \ref{fig6}(b) yields a dimensionless width $\delta E=0.40\pm0.04$. By comparing this value to one-particle Green's function calculations presented in Ref.~\cite{bostwick2010}, we deduce in turn a coupling constant $\alpha_G\approx 0.4$ for graphene/2D-Ag/SiC. Such a high value is well in line with Au-intercalated graphene/SiC which hosts a similar, semiconducting interlayer band structure \cite{forti2020}. In general, extrinsic dielectric screening of plasmon coupling in graphene will be rather poor if the substrate cannot provide any significant metallic density of states. The substantial warping of the present Dirac cone therefore underpins the truly semiconducting nature of 2D confined Ag with an estimated substrate dielectric constant \cite{bostwick2010,walter2011} $\epsilon_\mathrm{Ag}\approx4.4/\alpha_G-1=10.0\pm0.5$. Note that the residual screening is still stronger as compared to ($n$ doped) H-intercalated quasi-free-standing monolayer graphene with $\epsilon_\mathrm{H}=7.8$ \cite{bostwick2010}. We attribute this again to a small number of dangling bonds left behind at the SiC surface after Ag intercalation (already invoked in the discussion of Fig.\ \ref{fig4} above). Considering for instance the weak plasmon coupling in epitaxial monolayer graphene on SiC \cite{[{The small apparent elongation of the Dirac crossing for monolayer graphene/SiC can alternatively be explained by velocity enhancements of the linear $\pi$ bands in this 'anomalous region' combined with lifetime/experimental resolution broadening, see }] pramanik2022}, where Si dangling bonds are also present, the latter seem to be quite polarizable \cite{bostwick2010,walter2011} and could thus contribute to the dielectric screening in the present system.

For a given value of $\alpha_G$, the plasmaron-related renormalizations of the spectral function should scale linearly with the Dirac-point energy $E_D$ over a wide range of doping levels \cite{bostwick2010}. The diamond width $\delta E$ is thus expected constant as a function of $n$-type carrier density when plotted on the dimensionless energy axis scaled to the upper crossing point $E_0$. However, the dispersion does change drastically after K dosing as evident from the energy-momentum cut in Fig.\ \ref{fig6}(c) where the two crossing points at $\overline{\mathrm{K}}_\mathrm{gr}$ have essentially collapsed back into a singular Dirac point (cf.\ inset). The corresponding EDC in Fig.\ \ref{fig6}(d) exhibits a single (slightly asymmetric) peak which is significantly narrower as compared to the lineshape of the pristine system. A fit (now including a third peak to account for the spectral asymmetry) is dominated by the $E_0$ component identified at an absolute binding energy of $1.24$ eV \cite{Note1} and essentially limits $\alpha_G$ of the $n$-type doped system to below $\approx 0.1$. We attribute this massively enhanced screening to the ordered potassium overlayer which provides a metallic density of states through its parabolic K $4s$ band at $\overline{\Gamma}$, thereby acting like a switch for plasmaron quasiparticle formation in graphene. Significant screening just via the bare polarizability of ionized K atoms can be ruled out, since a previous study employing disordered potassium adatoms found no change in electron-plasmon coupling strength with enhanced $n$-type doping \cite{walter2011}.

Regarding the putative interlayer many-body interactions in the composite quantum system of 2D-Ag and graphene \cite{lee2022}, the potential of surface charge-transfer doping through potassium adsorption relies not only on donating additional carriers to the pristine electronic structure. As the partially ionized dopant system forms an ordered, metallic overlayer, the coupling of graphene's Dirac fermions to plasmons is efficiently suppressed. Enhanced dielectric screening of the graphene layer can hence stand as an alternative pathway to modify the interlayer proximity coupling and the apparently renormalized conduction band dispersion of 2D-Ag for which graphene plasmons are suggested to play an important role \cite{lee2022}.

Note that an alternative theory approach (involving electron correlations) technically reproduces the experimentally observed renormalizations of the Dirac cone as satellite features due to weak electron-plasmon interaction without finding proper plasmaron quasiparticles \cite{lischner2013}. Irrespective of whether the experimental data of Figs.\ \ref{fig6}(a) and (b) are interpreted in terms of weak \cite{lischner2013} or strong electron-plasmon coupling \cite{polini2008,bostwick2010,walter2011}, the resurgence of a single Dirac point upon introduction of the metallic (i.e., screening) potassium adlayer [Figs.\ \ref{fig6}(c) and (d)] is well consistent with both theoretical viewpoints.

As highlighted by the red arrows in Fig.\ \ref{fig6}(c) and the spectral derivative in the inset, an additional kink emerges in the $\pi$ band after potassium deposition. It is located at $\approx 1.7$ eV down from $E_F$, well below the Dirac point, and corresponds to an increase in band velocity by about a factor of $2$. Somewhat similar velocity enhancements have previously been reported near charge neutrality only ($n\approx10^{10}$ cm$^{-2}$), where they can be attributed to emerging electron-electron interactions \cite{elias2011,siegel2011}. At $n\approx10^{14}$ cm$^{-2}$ however, graphene offers a fundamentally different landscape for such many-body interactions, precluding any direct analogy. There is further no evidence for significant contributions from silver or potassium to the concerned region of energy-momentum space. Simple hybridization is therefore unlikely to produce the apparent kink in the $\pi$ band, whose origin still remains unclear.

\section{Conclusion}
\label{sec:concl}
In summary, we demonstrate how the band alignment in a vdW heterostack of epitaxial graphene and 2D confined silver can be changed via surface charge-transfer doping from an ordered $2\times2$ potassium overlayer. The Fermi level turns out to be moderately pinned inside the semiconducting gap of the Ag monolayer, whose conduction band minimum is not yet reached despite substantial electron transfer onto the heterostack ($\approx 10^{14}$ cm$^{-2}$). Yet surface charge-transfer doping manipulates the electronic structure of the heterostack beyond a mere band-structure realignment. While we obtain a trivial increase in graphene's electron-phonon coupling strength with enhanced $n$-type carrier density, plasmon-related renormalizations in the vicinity of the Dirac point are found to be largely suppressed. This can be attributed to enhanced screening via the emerging free-electron band structure of the ordered potassium overlayer.

Our high-resolution ARPES data further reveal an intriguing splitting of the saddle-point VHS of 2D confined Ag which is absent from the available calculations for this system \cite{wang2020,lee2022,hsu2012}. In addition, a high-energy kink emerges in the Dirac cone of graphene after surface charge-transfer doping. While being unique to the present sample system, dedicated studies will have to clarify the microscopic origin of these features.

Finally, we point out that the intercalation of bi- or multilayers of Ag beneath epitaxial graphene could be an alternative future pathway to modify the electronic structure of the intercalant (potentially closing the semiconducting gap) and thereby act on the combined heterostack, in analogy to the case of Au \cite{forti2020}.

\begin{acknowledgments}
We would like to thank Helmholtz-Zentrum Berlin for the allocation of synchrotron radiation beamtime under proposals 201-09081-ST and 202-09595-ST/R. Funding by the Deutsche Forschungsgemeinschaft (DFG) through Sta315/9-1 is gratefully acknowledged.
\end{acknowledgments}

\end{document}